\documentclass[a4paper,11pt]{article}

\usepackage{jcappub} 

\usepackage[T1]{fontenc} 
\usepackage{graphicx}
\usepackage{dcolumn}
\usepackage{bm}
\usepackage{amsmath}
\usepackage{amssymb}
\usepackage{hyperref}

\makeatletter

\title{\boldmath $\delta N$ formalism with gradient interactions}

\author[a]{S. Mohammad Ahmadi}
\author[a]{and Nahid Ahmadi}

\affiliation[a]{Department of Physics, University of Tehran, Karegar Ave North, Tehran 14395-547, Iran}

\emailAdd{mohammadahmadi@ut.ac.ir}
\emailAdd{nahmadi@ut.ac.ir}

\abstract{
	The standard $\delta N$ formalism is a cornerstone technique for calculating curvature perturbations on super-Hubble scales. However, its validity relies heavily on the separate universe assumption, in which spatial gradients are neglected. This approximation is known to break down in scenarios that are critical for primordial black hole formation, such as transitions to an ultra-slow-roll phase, where gradient interactions induce a significant non-conservation of the comoving curvature perturbation. In this paper, we introduce a framework for incorporating gradient corrections into the $\delta N$ formalism by adding an effective source term to the background Klein–Gordon equation. While preserving the nonlinear separate-universe dynamics at zeroth order, this approach consistently incorporates the leading gradient-sensitive effects and thereby improves the treatment of the nonlinear evolution of curvature perturbations up to the end of inflation, given initial conditions specified at horizon exit. By computing the equilateral non-Gaussianity parameter $f_{\mathrm{NL}}^{\mathrm{eq}}$, we demonstrate that our method captures some essential physical features missed by the standard $\delta N$ approach, offering a simple pathway to determine the nonlinear evolution expected from cosmological perturbation theory.
}

\keywords{inflation, cosmological perturbation theory, primordial non-Gaussianity, primordial black holes}

\begin{document}
\maketitle
\flushbottom

\section{Introduction}
\label{sec:intro}

In the context of single-field inflationary models, the scalar sector of cosmological perturbations is most effectively described by the gauge-invariant comoving curvature perturbation, denoted by $\mathcal{R}$. The dynamical evolution of this quantity serves as the fundamental link connecting vacuum quantum fluctuations generated during the inflationary epoch~\cite{starobinsky1980new, sato1981first, guth1981inflationary, linde1982new, albrecht1982reheating, lyth1999particle} to the late-time observables, such as the anisotropies in the cosmic microwave background and the large-scale structure of the universe. Beyond linear statistics, the properties of $\mathcal{R}$ are particularly crucial for understanding rare, large-amplitude fluctuations. These extreme events are of significant interest as they may have collapsed to form Primordial Black Holes (PBHs), which remain a compelling candidate for Dark Matter~\cite{zel1966hypothesis, hawking1971gravitationally, carr1974black, carr1975primordial, chapline1975cosmological, carr1984can}.

A powerful and widely used tool for studying the nonlinear evolution of the curvature perturbation is the $\delta N$ formalism~\cite{starobinsky1982dynamics, starobinskiǐ1985multicomponent, sasaki1996general, sasaki1998super, lyth2005general}. This formalism describes the dynamics of perturbations on super-Hubble scales and corresponds mathematically to the leading order of a spatial gradient expansion, parameterized by $\sigma = k/(aH) \ll 1$ (the ratio of the Hubble radius to the perturbation length scale). Physically, the $\delta N$ formalism relies on the Separate Universe Approach (SUA)~\cite{wands2000new, lyth2003conserved, rigopoulos2003separate}. In this picture, the universe is smoothed on scales larger than the horizon, such that each Hubble-sized region evolves independently as a homogeneous and isotropic FLRW spacetime determined solely by local field values. Essentially, the SUA treats the coarse-grained patches as uncoupled universes, ignoring the causal connections usually mediated by spatial gradients.

At the leading order in the gradient expansion, the governing perturbation equations simplify remarkably, coinciding with those of the homogeneous FLRW background once appropriate variable identifications are made~\cite{pattison2019stochastic}. Specifically, the background cosmic time is identified with the proper time along comoving trajectories, and the background scale factor corresponds to the local scale factor of the patch. These local values represent coarse-grained quantities averaged over horizon-sized separate universes.

The applicability of the SUA, and consequently the standard $\delta N$ formalism, is rooted in the assumption that spatial gradients can be safely neglected. Because of this, initial conditions for the separate universes are typically set several $e$-folds after Hubble crossing, allowing time for the decaying modes and gradient terms to vanish or when the modes are assumed to behave classically. It seems that working sufficiently far from the horizon crossing scale is a safe choice for matching with perturbation theory evolution. However, the results achieved by this approach are known to fail in several well‑motivated scenarios relevant to PBH formation. These include inflationary models featuring sudden or smooth transitions to an Ultra-Slow-Roll (USR) phase~\cite{jackson2024separate, briaud2025stochastic}. In such regimes, the rapid deceleration of the inflaton field can lead to the growth of non-adiabatic pressure perturbations and the persistence of super-Hubble gradient interactions. This results in the non-conservation of $\mathcal{R}$ on superhorizon scales, thereby invalidating the predictions of the standard $\delta N$ formalism over a finite range of super‑Hubble scales around the transition.

Recently, significant progress has been made in addressing the inherent limitations of the SUA. The authors of \cite{jackson2024separate} show that the full evolution in the super-Hubble regime can be constructed by matching fluctuations in the homogeneous field to the mode functions at (or shortly after) horizon crossing, provided that $k^2$ corrections are included at the matching scale. Although this approach goes beyond the SUA, it remains restricted to linear order in perturbation theory. In another approach, which allows for nonlinearities, it has been shown that the leading adiabatic gradient correction, of order $\mathcal{O}(k^2)$, can be encoded in the spatial curvature of each locally homogeneous FLRW slice~\cite{artigas2025extended}. In this approach, gradient interactions are applied only to the energy constraint.

Motivated by these developments, we propose a framework to incorporate gradient corrections systematically into the dynamics of superhorizon modes within a controlled gradient expansion. The effective source term that we introduce in the background Klein--Gordon equation is constructed by matching the separate-universe evolution to the gradient-expanded result of linear perturbation theory. In this sense, the source term itself is fixed at linear order in perturbations. However, once included in the background evolution, its effect is propagated through the nonlinear $\delta N$ mapping. Within the regime of validity of the gradient expansion, this construction is equivalent to the corresponding gradient-expanded linear perturbation theory result, while retaining the simplicity and nonlinear separate-universe structure of the $\delta N$ formalism. It is also important to note that this construction is effective only within the domain of validity of the SUA, which is dictated by the condition involving the first and second slow-roll parameters, $\epsilon \ll 1$ \cite{cruces2023update}, and we restrict ourselves to spatial gradients at leading order in perturbation theory.

The key quantity that captures the leading nonlinear and non-Gaussian effects arising from gradient interactions is the bispectrum. This statistic plays a crucial role in determining the probability distribution of perturbations and, by extension, the PBH abundance~\cite{byrnes2012primordial, young2013primordial, passaglia2019primordial, atal2019role, biagetti2021formation, taoso2021non, young2022peaks, ferrante2022primordial, gow2023non}, which is exponentially sensitive to non-Gaussian tails. Accurate determination of the bispectrum requires a precise treatment of second-order perturbations. Since the $\delta N$ formalism is intrinsically non-perturbative, it naturally captures the nonlinear evolution of superhorizon fluctuations. By extending it to include gradients, the shape and amplitude of intrinsic non-Gaussianity~\cite{komatsu2001acoustic, maldacena2003non, bartolo2004non, liguori2010primordial, chen2010primordial, wang2014inflation, cai2018revisiting, namjoo2025geometry} can be analyzed within the $\delta N$ framework more rigorously. In this work, we compute the equilateral non-Gaussianity parameter $f_{\mathrm{NL}}^{\mathrm{eq}}$ for two illustrative models. A comparison of our results with those of the standard $\delta N$ approach, as well as with results based on the standard in-in formalism, highlights the necessity of our proposed framework.

This paper is organized as follows. In section~\ref{sec:SUA}, we review the SUA and its implementation through the higher-order matching method and the standard $\delta N$ formalism. We emphasize the role of spatial gradients and discuss the limitations inherent in these approaches. Section~\ref{sec:Incorporating_gradient} is devoted to the development of our source-term approach. In subsection~\ref{sec:the_source_term} we introduce a gradient-corrected $\delta N$ framework by incorporating an effective source term into the background Klein--Gordon equation and demonstrate its formal equivalence to the gradient expansion. We then validate the formalism by comparing its predictions with linear perturbation theory for two representative inflationary models in subsection~\ref{sec:LPT}. As an application, we compute the equilateral non-Gaussianity parameter and show that gradient interactions play a crucial role, comparing our results with those obtained using the standard $\delta N$ approach in subsection~\ref{sec:NG}. Section~\ref{sec:Summary} summarizes our findings. In the appendix, we review the $\delta N$ formalism with spatial curvature introduced in Ref.~\cite{artigas2025extended} and compare its results with those of our proposed $\delta N$ formalism.

\section{\boldmath Separate universe approach and $\delta N$ formalism}

In this section, we review the SUA, its formulation within linear cosmological perturbation theory, and its practical implementation through the $\delta N$ formalism. Particular emphasis is placed on the role of spatial gradient interactions, which are systematically neglected both in the SUA and in the standard $\delta N$ framework. We further discuss methods that incorporate gradient effects via an expansion.

\subsection{Separate universe approach}
\label{sec:SUA}

For a four-dimensional metric tensor with components $g_{\mu\nu}$ minimally coupled to a scalar field $\phi$, the Einstein--Hilbert action is
\begin{equation}
	S = \int \mathrm{d}^4 x \sqrt{-g} \left[ \frac{M_{\text{Pl}}^{2}}{2} R 
	- \frac{1}{2} g^{\mu \nu} \partial_{\mu} \phi \, \partial_{\nu} \phi 
	- V(\phi) \right].
	\label{action}
\end{equation}
If we restrict our analysis to the linearized version of cosmological perturbation theory, the most general form (without making any gauge choice) of the line element with perturbations around an FLRW background is
\begin{equation}
	\mathrm{d}s^{2}
	= -(1 + 2A)\,\mathrm{d}t^{2}
	+ 2a\,\partial_i B\,\mathrm{d}x^{i}\mathrm{d}t
	+ a^{2}\left[(1 - 2\psi)\delta_{ij}
	+ 2\,\partial_i \partial_j E \right]
	\mathrm{d}x^{i}\mathrm{d}x^{j},
	\label{eq:metric_scalar}
\end{equation}
where $A$, $B$, $\psi$, and $E$ are scalar metric perturbations.

This decomposition makes explicit that
inhomogeneities enter through spatial gradient terms
$(\partial_i)$.
On sufficiently large scales $(k \rightarrow 0)$, where
these gradients can be neglected, each local Hubble
patch evolves as an effectively homogeneous and
isotropic universe, up to small perturbations of the
background quantities. This observation underlies the SUA, in which super-Hubble perturbations can be
absorbed into locally perturbed background variables.
Consequently, the dynamics of a canonical scalar field
$\phi(t)$ in each patch are governed by the homogeneous
Klein--Gordon (KG) equation,
\begin{equation}
	\ddot{\phi} + 3H\dot{\phi} + V_{,\phi} = 0,
	\label{eq:KG_background}
\end{equation}
where overdots denote derivatives with respect to cosmic
time $t$.

To justify this formally, let us consider the evolution of the fluctuations. The equation of motion for the scalar field follows from varying the action~\eqref{action} with respect to $\phi$, which yields the covariant KG equation
\begin{equation}
	\Box \phi - V_{,\phi} = 0 ,
	\label{eq:KG_covariant}
\end{equation}
where the d'Alembertian operator $\Box \equiv g^{\mu\nu}\nabla_\mu\nabla_\nu$ is constructed from the perturbed metric given in Eq.~\eqref{eq:metric_scalar}. Expanding Eq.~\eqref{eq:KG_covariant} and retaining terms up to linear order in perturbations, one obtains the KG equation for the Fourier modes of the field fluctuation,
\begin{equation}
	\ddot{\delta\phi}_k
	+ 3H \dot{\delta\phi}_k
	+ \left(\frac{k^2}{a^2} + V_{,\phi\phi}\right)\delta\phi_k
	=
	-2 V_{,\phi} A_k
	+ \dot{\phi}
	\left(
	\dot{A}_k + 3\dot{\psi}_k
	+ \frac{k^2}{a^2}\left[a^2\dot{E}_k - a B_k\right]
	\right),
	\label{eq:kg_perturbed}
\end{equation}
where the background equation of motion for $\phi$ has been used to simplify the result. The metric perturbations entering the right-hand side are constrained by the Einstein field equations and are not independent. In particular, the energy and momentum constraint equations take the form
\begin{align}
	3H\left(\dot{\psi}_k + H A_k\right)
	+ \frac{k^2}{a^2}
	\left[
	\psi_k + H\left(a^2\dot{E}_k - a B_k\right)
	\right]
	&=
	-\frac{1}{2M_{\rm Pl}^2}
	\left[
	\dot{\phi}\left(\dot{\delta\phi}_k - \dot{\phi} A_k\right)
	+ V_{,\phi}\delta\phi_k
	\right],
	\label{eq:energy_constraint}
	\\
	\dot{\psi}_k + H A_k
	&=
	\frac{\dot{\phi}}{2M_{\rm Pl}^2}\,\delta\phi_k .
	\label{eq:momentum_constraint}
\end{align}

It is convenient to combine the field and metric perturbations into a single gauge-invariant quantity, the Sasaki--Mukhanov (MS) variable, defined as
$Q_k \equiv \delta\phi_k + \frac{\dot{\phi}}{H}\,\psi_k$.
Using the constraint equations to eliminate the metric perturbations from the perturbed KG equation, the dynamics can be written entirely in terms of $Q_k$, leading to~\cite{pattison2019stochastic}
\begin{equation}
	\ddot{Q}_k
	+ 3H \dot{Q}_k
	+ \left(
	\frac{k^2}{a^2}
	+ V_{,\phi\phi}
	- \frac{1}{M_{\rm Pl}^2 a^3}
	\frac{d}{dt}\!\left(\frac{a^3}{H}\dot{\phi}^2\right)
	\right) Q_k
	= 0 .
	\label{perturbative_equation_deltaphi}
\end{equation}
This equation describes the linear evolution of scalar perturbations and is valid in all gauges. For the purposes of this section, it can be compared with its counterpart in SUA at the linear level either in gauge-invariant form or in a fixed gauge. The linearly perturbed form of the homogeneous and isotropic background equation of motion \eqref{eq:KG_background} under the mapping $\phi \rightarrow \phi + \delta\phi$, $H \rightarrow H + \delta H$, and $dt \rightarrow (1+A)\,dt$ is given by~\cite{pattison2019stochastic}
\begin{equation}
	\ddot{\delta\phi} + \left(3H + \frac{\dot{\phi}^2}{2M_{\rm Pl}^2H}\right)\dot{\delta\phi} \: + \left(\frac{\dot{\phi}}{2M_{\rm Pl}^2H}V_{,\phi} + V_{,\phi\phi}\right)\delta\phi
	- \dot{\phi}\dot{A} - \left(2\ddot{\phi}+3H\dot{\phi}+ \frac{\dot{\phi}^3}{2M_{\rm Pl}^2H}\right)A = 0 \, 
\end{equation}
Specializing to the spatially flat (\(\psi = B = 0\)) or uniform expansion (\(\delta N = B = 0\)) gauge, the above expression can be written as \cite{pattison2019stochastic}
\begin{equation}
	\ddot{Q}_k
	+ 3H \dot{Q}_k
	+ \left(
	V_{,\phi\phi}
	- \frac{1}{M_{\rm Pl}^2 a^3}
	\frac{d}{dt}\!\left(\frac{a^3}{H}\dot{\phi}^2\right)
	\right)Q_k
	= 0 \,.
	\label{pertubed_BG_eqaation}
\end{equation}

Comparing Eqs.~\eqref{perturbative_equation_deltaphi} and \eqref{pertubed_BG_eqaation}, the SUA appears to reproduce the results of linear perturbation theory in the $k \to 0$ limit. However, as emphasized in foundational works on the $\delta N$ formalism~\cite{kodama1998evolution, sasaki1998super}, deriving perturbation equations solely from the homogeneous background dynamics omits terms dictated by the momentum constraint. In the derivation of Eq.~\eqref{pertubed_BG_eqaation}, this constraint is effectively imported from perturbation theory. If one instead works strictly with perturbed background equations while omitting the momentum constraint, the evolution equation for the gauge-invariant MS variable acquires additional terms of order $\mathcal{O}(\epsilon\eta)$, which manifest as an additional decaying mode proportional to $\dot{\phi}\int dt/a^{3}$ in the solution~\cite{cruces2022review}, provided one assumes a constant-roll phase for inflation \cite{martin2013ultra, motohashi2015inflation, odintsov2017inflation, anguelova2018systematics, morse2018large, yi2018constant, lin2019dynamical, ghersi2019observational, mohammadi2023constant, mohammad2024analytical}.
Therefore, the long-wavelength equivalence between SUA and linear perturbation theory holds precisely only in the small-$\epsilon$ limit.  
Deviations from this equivalence appear in scenarios such as punctuated inflation~\cite{jain2008double, jain2010tensor, hazra2013bingo}, though analyzing such cases lies beyond the scope of this work.

Working in phase space and reducing to homogeneous and isotropic degrees of freedom, the authors of \cite{artigas2022hamiltonian} showed that the resulting perturbed dynamical system for the scenario described above reproduces the predictions of cosmological perturbation theory at leading order, either at the level of the Hamiltonian or at the level of the dynamical equations. This result implies that (within the domain of validity of $\epsilon \eta \ll 1$) the SUA provides an accurate description of large-scale fluctuations.

\subsection{Gradient expansion}

The SUA corresponds to the leading-order term in a spatial gradient expansion. At linear order in perturbation theory, the full dynamics of the comoving curvature perturbation $\mathcal{R}_k$ are governed by the MS equation. This equation can be obtained by expressing the variable $Q$ in Eq.~\eqref{perturbative_equation_deltaphi} in terms of the gauge-invariant curvature perturbation via the relation $Q_k = \dot{\phi} \mathcal{R}_k /H$, which leads to
\begin{equation}
	\mathcal{R}''_k + 2 \frac{z'}{z}\, \mathcal{R}'_k + k^2 \mathcal{R}_k = 0 ,
	\label{eq:MS_for_R}
\end{equation}
where primes denote derivatives with respect to conformal time $\eta$, and $z \equiv a\dot{\phi}/H$. A formal solution to this differential equation can be constructed as an expansion in powers of $k^2$~\cite{jackson2024separate}. By decomposing $\mathcal{R}_k$ into linearly independent growing and decaying modes,
\begin{equation}
	\mathcal{R}_k(\eta) = \mathcal{G}(\eta) + \mathcal{D}(\eta) \,,
	\label{gradient_expansion_base_formula}
\end{equation}
each mode admits a power-series expansion in the wavenumber:
\begin{equation}
	\mathcal{G}(\eta) = \sum_{i=0}^{\infty} \mathcal{G}_i(\eta)\, k^{2i} \,,
	\qquad
	\mathcal{D}(\eta) = \sum_{i=0}^{\infty} \mathcal{D}_i(\eta)\, k^{2i} \,.
	\label{ad_nad_expansion}
\end{equation}
The homogeneous solutions at leading order (the $k \to 0$ limit) are
\begin{equation}
	\mathcal{G}_0(\eta) \equiv C_k \,,
	\qquad
	\mathcal{D}_0(\eta) \equiv D_k
	\int_{\eta}^{\eta_{\rm ref}}
	\frac{d\tilde{\eta}}{z^2(\tilde{\eta})} \,,
	\label{zeroth_order_gradients}
\end{equation}
which are the solutions modeled by the SUA. Here, $C_k$ and $D_k$ are integration constants fixed by initial conditions, and $\eta_{\rm ref}$ is an arbitrary reference time. Higher-order corrections can be generated recursively. The coefficients $\mathcal{D}_i(\eta)$ (and similarly for $\mathcal{G}_i$) satisfy recursive integral relations of the form:
\begin{equation}
	\mathcal{D}_i(\eta)
	= -
	\int_{\eta_{\rm ref}}^{\eta}
	\frac{d\eta_1}{z^2(\eta_1)}
	\int_{\eta_{\rm ref}}^{\eta_1}
	d\eta_2\,
	z^2(\eta_2)\, \mathcal{D}_{i-1}(\eta_2) \,.
	\label{eq:recursive_Gn}
\end{equation}
The SUA corresponds to truncating this expansion at zeroth order. However, gradient corrections ($i \geq 1$) become essential when the full evolution of the curvature perturbation on super-Hubble scales is desired. This occurs in non-attractor models or during sharp transitions in the inflationary potential. The remarkable feature of these terms is discussed in detail in Ref.~\cite{briaud2025stochastic}.

\subsection{Homogeneous and higher-order matching methods}

In practical applications, the SUA is often implemented via a \textit{homogeneous matching} procedure~\cite{jackson2024separate}. This technique involves matching the solution obtained from full linear perturbation theory (starting with Bunch–Davies initial conditions and containing all orders in $k^2$) below a certain scale, to the homogeneous part of the gradient-expansion series \eqref{zeroth_order_gradients} above that scale. This matching allows one to bridge the quantum regime with the classical separate universe evolution.

The matching time $\eta_*$ is defined by the condition $k = \sigma a(\eta_*) H(\eta_*)$. Typically, the parameter $\sigma$ is chosen to be smaller than one. This choice guarantees the smallness of the expansion parameter at the matching surface, ensuring that gradient terms are subdominant and thereby establishing the validity of the SUA.

Requiring continuity of $\mathcal{R}_k$ and its derivative $\mathcal{R}'_k$ at $\eta_*$ fixes the integration constants in Eq.~\eqref{zeroth_order_gradients}:
\begin{equation}
	\hat{C}_k
	=
	\mathcal{R}_{k*}
	+
	z_*^2 \mathcal{R}'_{k*}
	\int_{\eta_*}^{0}
	\frac{d\tilde{\eta}}{z^2(\tilde{\eta})} \,,
	\qquad
	\hat{D}_k
	=
	-\,z_*^2 \mathcal{R}'_{k*} \,.
\end{equation}
Following the notation of Ref.~\cite{jackson2024separate}, starred quantities are evaluated at the matching time, and those with hats denote values obtained via homogeneous matching.

Although this method is valid in both slow-roll and standard non–slow-roll regimes, it fails over a finite range of scales in models with an enhanced power spectrum, precisely because the gradient terms are not negligible during the enhancement phase. To address this, one can perform a \textit{higher-order matching} procedure. Instead of truncating the expansion at zeroth order, the authors of Ref.~\cite{jackson2024separate} suggest adding the next-to-leading order term in the gradient expansion to $\mathcal{R}_k$ for $\eta > \eta_*$. By including a sufficient number of higher-order terms, this method ensures a smoother transition from the full quantum mode evolution at the matching time to the gradient-expanded solution. Based on an inductive generalization, one may then employ the full gradient-expansion series,
\begin{equation}
	\hat{\mathcal{R}}_k(\eta)
	= \mathcal{R}_{k*}\,u_{\rm ad}(\eta)
	+ \mathcal{R}'_{k*}\,u_{\rm nad}(\eta) \,,
	\label{higher_order_matching}
\end{equation}
and apply this to find the superhorizon evolution of perturbations beyond the slow-roll approximation and the SUA. Here, the adiabatic ($u_{\rm ad}$) and non-adiabatic ($u_{\rm nad}$) mode functions are obtained order by order from Eq.~\eqref{eq:recursive_Gn} and its analogous expression for $\mathcal{G}_i$, respectively.

\subsection{$\delta N$ formalism}

Another powerful implementation of the SUA is provided by the $\delta N$ formalism~\cite{starobinsky1982dynamics, starobinskiǐ1985multicomponent, sasaki1996general, sasaki1998super, lyth2005general}.
On sufficiently large scales, the curvature perturbation on comoving hypersurfaces can be identified with the perturbation in the local expansion history, $\delta N$. In this framework, all information about inhomogeneities is encoded in the perturbed initial conditions of an ensemble of locally homogeneous background patches which then evolve nonlinearly.

In practice, one computes the difference between the number of $e$-folds realized along a locally perturbed trajectory—obtained by evolving the background equations from the initial conditions $\phi_* + \delta\phi_{k*}$ and $\dot{\phi}_* + \delta\dot{\phi}_{k*}$ up to a final comoving hypersurface (usually at the end of inflation)—and the background number of $e$-folds:
\begin{equation}
	-\mathcal{R}_k \simeq \delta N_k
	=
	N(\phi_* + \delta\phi_{k*}, \dot{\phi}_* + \delta\dot{\phi}_{k*})
	- N(\phi_*, \dot{\phi}_*) \,.
	\label{deltaN_formula}
\end{equation}
The final comoving hypersurface should be chosen sufficiently late during inflation so that the evolution has settled back into the slow-roll attractor regime, where the approximation $ \delta N_k \simeq - \mathcal{R}_k$ is reliable.\footnote{The comoving curvature perturbation $\mathcal{R}_k$ is generically different from the curvature perturbation on uniform-density hypersurfaces, $\zeta_k$, but they can be related via non-local terms. Although these terms are negligible in the slow-roll regime, the difference can be exponentially large in other regimes such as USR inflation \cite{cruces2023update}. Since the final hypersurface for the $\delta N$ evaluation is chosen to be sufficiently late, in slow-roll regimes the identity $\zeta_k = -\mathcal{R}_k$ holds reliably.
} Expanding \eqref{deltaN_formula} yields
\begin{align}
	\delta N_k = N_a \, \delta X_{a,k}
	+ \frac{1}{2} N_{ab} \int \frac{d^3q}{(2 \pi)^3} \delta X_{a,q} \delta X_{b,k-q}
	+ \dots,
	\label{expand_R_second}
\end{align}
which effectively represents a matching prescription analogous to homogeneous matching. The coefficients of this expansion are purely kinematic quantities derived from the background evolution. Here, repeated indices are summed over, $X_a$ denotes the $a$-th component of the phase-space vector $X = ( \phi_{\mathrm{in}}, \dot{\phi}_{\mathrm{in}})$, and derivatives are defined as
\begin{equation}
	N_a \equiv \left. \frac{\partial N}{\partial X_a} \right|_{\eta_*},
	\qquad
	N_{ab} \equiv \left. \frac{\partial^2 N}{\partial X_a \partial X_b} \right|_{\eta_*}.
\end{equation}

In Ref.~\cite{artigas2022hamiltonian}, it was shown that the perturbed dynamical equations obtained in the SUA match the results of linear order perturbation theory (in which the anisotropic degrees of freedom are set to zero) when sufficiently large scales are considered. This proves that the homogeneous matching procedure and the linearized $\delta N$ formalism (which relies on the SUA above a matching scale) are equivalent.
Although these approximate formalisms are valid in most situations, their inability to account for gradient interactions constitutes a major limitation in certain cases of interest. In principle, the higher-order matching method can be used to include such interactions; however, in doing so, the nonlinearities inherent in the fully non-perturbative theory are sacrificed in favor of gradient terms.

The $\delta N$ formalism and the higher-order matching methods each come with approximations that simplify the complex dynamics of cosmological systems. The primary strength of the $\delta N$ formalism is its ability to extend beyond linear perturbation theory. By expanding $N$ to higher orders in $\delta \phi_{k*}$ and $\delta\dot{\phi}_{k*}$, it establishes a general framework for computing nonlinear curvature perturbations and non-Gaussianities, whereas the linear matching method lacks this capability~\cite{yokoyama2008primordial, wands2010local, hooshangi2022rare, hooshangi2023tail}. Nevertheless, the standard $\delta N$ formalism is inherently a zeroth-order approximation in the gradient expansion. Unlike the higher-order matching procedure, which can incorporate finite-$k$ corrections to account for the effects of spatial gradients, the standard $\delta N$ formalism relies strictly on the SUA. Consequently, the method breaks down when gradient interactions become significant~\cite{jackson2024separate, briaud2025stochastic}. These limitations motivate the development of an extended $\delta N$ formalism that goes beyond the long-wavelength approximation and consistently incorporates all gradient effects. In the following, we show that it is possible to generalize the $\delta N$ formalism by adopting a full linear treatment of gradient interactions as the sole assumption.

\section{\boldmath Incorporating gradient terms into the $\delta N$ formalism}
\label{sec:Incorporating_gradient}

In the previous section, we presented a schematic picture of how one might obtain the fluctuation dynamics from the background equations of motion. In this section, we demonstrate how gradient corrections can be incorporated into the $\delta N$ formalism, thereby extending its applicability. We will then use this framework to compare the $\delta N$ formalism with linear perturbation theory and to perform a full numerical analysis of deviations from Gaussian statistics.

\subsection{Source term of gradient interactions}
\label{sec:the_source_term}
A heuristic approach is proposed in Ref.~\cite{artigas2025extended} to extend the $\delta N$ formalism by accounting for suppressed gradient terms. That work focuses solely on the $\mathcal{O}(k^2)$ contribution to the energy constraint, while gradient interactions in the KG equation were neglected. In this way, the effect of adiabatic $k^2$ gradient corrections was encoded into the spatial curvature of each locally homogeneous, expanding FLRW patch. In Appendix~\ref{app:review_extended_deltaN}, we review this approach and its impact on the power spectrum.

In our approach to extend the applicability of the $\delta N$ formalism, we introduce a source term in the field equation of motion. To derive the particular source term that correctly captures the effects of gradient interactions, we compare the perturbed KG equation~\eqref{perturbative_equation_deltaphi} with the perturbed background equation~\eqref{pertubed_BG_eqaation}. The failure of the SUA to capture gradient effects can be traced to the absence of the $k^2$ term responsible for spatial gradients in the perturbed background equation. To bridge this gap, we explicitly restore the missing $k^2 Q_k / a^2$ contribution in the background KG equation by introducing a source term. Using the relation $Q_k = \dot{\phi} \mathcal{R}_k/H$, the modified background equation takes the form
\begin{equation}
	\ddot{\phi} + 3H\dot{\phi} + V_{,\phi} = \mathcal{S}_k ,
	\qquad
	\mathcal{S}_k = - \frac{k^2 \dot{\phi}}{a^2 H}\, \hat{\mathcal{R}}_k ,
	\label{KG_with_correction}
\end{equation}
where we have moved to $k$-space. Note that $\hat{\mathcal{R}}_k$ is used in the source term instead of $\mathcal{R}_k$, since the evolution of the curvature perturbation is assumed to be known only up to the matching time, beyond which it is determined using the gradient expansion. Physically, the introduction of this source term acts as a proxy for the spatial Laplacian, forcing the background fields to respond to local curvature in a manner that mimics the full wave equation \eqref{eq:KG_covariant}.

The source term is defined in terms of the gauge-invariant dynamical scalar degree of freedom $\mathcal{R}_k$ in the isotropic sector of phase space, which ensures consistency with cosmological perturbation theory. By working in uniform expansion gauge, the anisotropic degrees of freedom decouple from the isotropic ones \cite{artigas2022hamiltonian}, and a source term originating from the isotropic sector cannot mix these two sectors in phase space.  Full consistency with perturbation theory is ensured by mapping the variable $Q$ in Eq.~\eqref{perturbative_equation_deltaphi} to the curvature perturbation via the relation $Q_k = \dot{\phi} \mathcal{R}_k / H$.

This source term lies at the heart of the higher-order matching method. The equivalence of the higher-order matching method to this new $\delta N$ formalism can be analytically demonstrated. Substituting $Q_k = z \mathcal{R}_k /a$ into the linearized form of the background equation~\eqref{KG_with_correction} yields $\left( z^2 \mathcal{R}'_k \right)' = - k^2 z^2 \hat{\mathcal{R}}_k$.
Using the matching relation \eqref{higher_order_matching} for $\mathcal{O}(k^{2(i+1)})$ terms, we have
\begin{equation}
	\left( z^2 \mathcal{R}^{(i+1)\prime}_k \right)'
	= - z^2 \bigl( \mathcal{G}_{i} + \mathcal{D}_{i} \bigr)\,.
\end{equation}
Integrating this equation twice yields
\begin{equation}
	\mathcal{R}^{(i+1)}_k(\eta)
	= - \int_{\eta_*}^{\eta} \frac{d\eta_1}{z^2(\eta_1)}
	\int_{\eta_*}^{\eta_1} d\eta_2\, z^2(\eta_2)\,
	\bigl( \mathcal{G}_{i}(\eta_2) + \mathcal{D}_{i}(\eta_2) \bigr)
	= \mathcal{G}_{i+1}(\eta) + \mathcal{D}_{i+1}(\eta)\,,
\end{equation}
where we have assumed $\mathcal{R}^{(i)}_{k*} = \mathcal{R}^{(i)\prime}_{k*} = 0$ for $i \geq 1$. The right-hand side of this relation corresponds precisely to the $(i+1)$th-order term in the gradient expansion, Eq.~\eqref{ad_nad_expansion}. This demonstrates a strong mathematical concurrence between the higher-order matching method and (linear) source-term method. However, we remind the reader that this entire discussion is valid only in the small-$\epsilon$ limit (see the discussion following Eq.~\eqref{pertubed_BG_eqaation}). Although the effect of the artificial decaying mode---which is suppressed by a small $\epsilon$---is negligible on the power spectrum, it might have a palpable effect on the non-Gaussianity (which we discuss later), since the non-Gaussian signal itself is $\mathcal{O}(\epsilon)$. Nevertheless, we will demonstrate using accurate numerical methods that this gradient-corrected $\delta N$ formalism is able to predict the main non-Gaussian features of interest. Despite slight inaccuracies, it constitutes a significant improvement over the standard $\delta N$ formalism.

The application of this method requires that the $\delta N$ expansion be carried out with respect to all integration constants, $\{\phi_{\mathrm{in}}, \dot{\phi}_{\mathrm{in}}, \mathcal{R}_{\mathrm{in}}, \mathcal{R}'_{\mathrm{in}}\}$. However, in single-clock inflation, only two of these quantities are truly independent degrees of freedom. In this work, we choose the independent parameters to be $X = (\phi_{\mathrm{in}}, \dot{\phi}_{\mathrm{in}})$, in direct analogy with the standard $\delta N$ formalism, and use the linear constraint relation $\mathcal{R}_k = H Q_k / \dot{\phi}$ to eliminate the remaining two.

\subsection{Comparison with linear perturbation theory}
\label{sec:LPT}

This section illustrates the reliability of the formalism developed above in predicting the results of linear perturbation theory. We base our analysis on two distinct inflationary models that exhibit nontrivial dynamics.

As a first example, consider a potential featuring a Gaussian bump, given by
\begin{equation}
	V(\phi) = 
	\frac{V_0 \phi^2}{m^2 + \phi^2}
	\left[
	1 + K \exp\left(
	- \frac{1}{2}\frac{(\phi - \phi_0)^2}{\Sigma^2}
	\right)
	\right],
	\label{eq:potential_bump}
\end{equation}
with the parameter values $V_0 = 0.143 M_{\mathrm{Pl}}^4$, $m = 0.5\,M_{\mathrm{Pl}}$, $K = 1.876 \times 10^{-3}$, $\phi_0 = 2.005\,M_{\mathrm{Pl}}$, $\Sigma = 1.993 \times 10^{-2}\,M_{\mathrm{Pl}}$, and $\phi_{in}= 2.605 M_{\mathrm{Pl}}$~\cite{mishra2020primordial}. This model exhibits a smooth transition into a transient USR phase, leading to a gradual enhancement of the curvature perturbations. As the second example, we examine Starobinsky's linear potential~\cite{starobinskij1992spectrum, martin2012scalar, martin2014sharp, ahmadi2022quantum}. This is a classic model featuring a sharp transition in the slope of the potential, which induces a temporary USR phase. This leads to a distinct enhancement of the power spectrum. The potential is defined piecewise as:
\begin{equation}
	V(\phi) =
	\begin{cases}
		V_0 + A_{+}(\phi - \phi_T), & \phi \geq \phi_T, \\[6pt]
		V_0 + A_{-}(\phi - \phi_T), & \phi < \phi_T.
	\end{cases}
	\label{eq:starobinsky_potential}
\end{equation}
We choose the parameter values $V_0 = 0.137 M_{\mathrm{Pl}}^4$, $A_+=4.56 \times 10^{-3} M_{\mathrm{Pl}}^3$, $A_-=1.139 \times 10^{-3} A_+$, $\phi_T=0$, and $\phi_{in}= 0.4 M_{\mathrm{Pl}}$. These parameter sets produce a power spectrum with an enhancement sufficient for the resulting asteroid-mass PBH abundance to account for the total dark matter content~\cite{cole2023primordial}. The large-scale power is normalized to unity for numerical convenience.

By definition, the standard $\delta N$ formalism applies to superhorizon scales beyond a certain length scale, normally commensurate with the Hubble radius. However, there is no general prescription for defining this scale in cosmological spacetimes. Agreement with the results of perturbation theory therefore justifies the choice of matching time. Here, we set $\sigma = 1$, corresponding to matching at the time of Hubble horizon crossing.\footnote{The requirement of a small $\sigma$ is a limitation specific to the standard $\delta N$ formalism, not to our proposed gradient-corrected formalism. In fact, one of the primary results of our work is that the gradient-corrected $\delta N$ formalism remains applicable even when a relatively large value of $\sigma$ is chosen. This is because the full non-adiabatic evolution of the curvature perturbation at linear order is incorporated within our framework.} At this scale, we incorporate gradient terms into the source term~\eqref{KG_with_correction} order by order and compare the results with the corresponding orders obtained using the higher-order matching method, for which full agreement is expected.

\begin{figure}[tbp]
	\centering
	\includegraphics[width=0.49\textwidth]{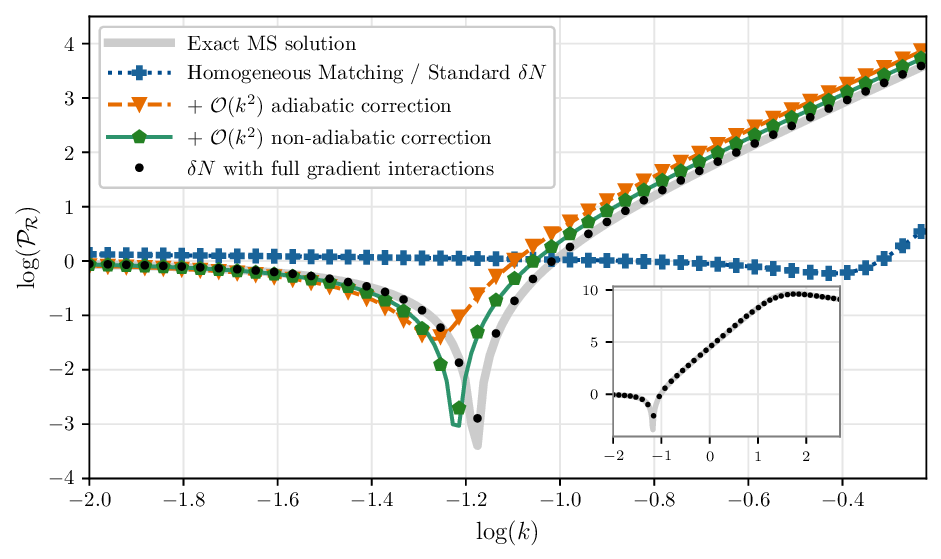}
	\hfill
	\includegraphics[width=0.49\textwidth]{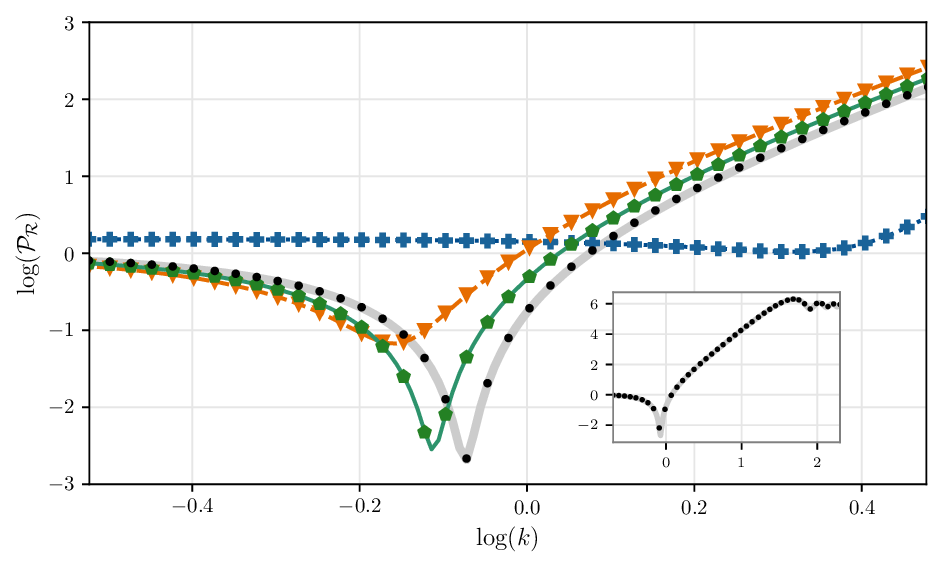}
	\caption{Comparison of the power spectra of the curvature perturbation, $\mathcal{P}_\mathcal{R} = \frac{k^3}{2 \pi^2} \left| \mathcal{R}_k \right|^2$ computed using the $\delta N$ formalism~\eqref{expand_R_second}, the higher-order matching method~\eqref{higher_order_matching}, and the exact MS equation~\eqref{eq:MS_for_R}, for the Gaussian bump (left panel) and Starobinsky (right panel) models \cite{github_code}. The gray solid line denotes the exact MS solution, while the colored lines represent the results obtained by step-by-step inclusion of $\mathcal{O}(k^2)$ corrections in the matching method. Colored markers indicate the corresponding $\delta N$ results. The black dot markers show the $\delta N$ results with full gradient interactions. The inset in each panel compares the MS solution with the $\delta N$ results including full gradients over a wider range of wavenumbers.}
	\label{fig:compare_HM_deltaN_bump}
\end{figure}

This equivalence is illustrated in Figure~\ref{fig:compare_HM_deltaN_bump}. The figure clearly shows that the fit to full perturbation theory improves significantly with the inclusion of higher-order gradient expansion terms. It is easy to see that non-adiabatic corrections are not negligible, highlighting the need to treat the non-adiabatic modes carefully. The role of these modes becomes clear when the MS equation is written in the form of a driven oscillator: $\mathcal{R}_k'' + 2 \frac{z'}{z} \mathcal{R}_k' + k^2 \mathcal{R}_k = \mathcal{F}_k(\eta)$. Comparing the homogeneous system (where $\mathcal{F}_k(\eta)=0$) with the driven one (where $\mathcal{F}_k(\eta) \sim \lambda\,\mathcal{R}_k^{\rm nad}$, with $\lambda \sim \mathcal{O}(k^2)$) reveals substantial differences. The former describes a system with a restoring force, while the latter is driven by an external source. In the presence of strong damping, $z'/z \gg \sqrt{\lambda}$, the system asymptotically approaches $\mathcal{R}_k^{\rm nad}$ at late times. This situation can arise during a sharp transition, where the secular growth of the non-adiabatic mode drives the system away from the simple attractor model ($\mathcal{F}_k = 0$).

\subsection{Non-Gaussianity}
\label{sec:NG}

Having established that the gradient-corrected formalism reproduces the gradient expansion of perturbation theory at linear order, we now use it to estimate the leading non-Gaussian signal through the momentum-dependent nonlinear parameter $f_{\rm NL}$. The correlator of three Fourier modes is given by
\begin{equation}
	\langle \mathcal{R}_{k_1} \mathcal{R}_{k_2} \mathcal{R}_{k_3} \rangle
	= \frac{9}{10} \frac{(2 \pi)^7 \delta^{(3)} (\mathbf{K})}{(k_1 k_2 k_3)^2}
	S(\mathbf{k}_1,\mathbf{k}_2,\mathbf{k}_3).
	\label{3_point_function}
\end{equation}
Here, $\textbf{K} = \mathbf{k}_1 + \mathbf{k}_2 + \mathbf{k}_3$ ensures momentum conservation, and the shape function $S(\mathbf{k}_1,\mathbf{k}_2,\mathbf{k}_3)$ is given by
\begin{equation}
	S(\mathbf{k}_1,\mathbf{k}_2,\mathbf{k}_3) = f_{\rm NL} \frac{k_3^3 \mathcal{P}_{\mathcal{R}}(k_1) \mathcal{P}_{\mathcal{R}}(k_2) + \text{2 perms}}{k_1k_2k_3}.
\end{equation}
The three-point function can also be computed using the $\delta N$ expansion~\eqref{expand_R_second}, yielding a similar expression but with the shape function defined via derivatives of $N$:
\begin{equation}
	S(\mathbf{k}_1,\mathbf{k}_2,\mathbf{k}_3) = \frac{5}{6}\frac{k_3^3 N_a N_b N_{cd} \mathcal{P}_{ac}(k_1) \mathcal{P}_{bd}(k_2) + \text{2 perms}}{k_1k_2k_3},
	\label{Shape_function_dN}
\end{equation}
where the cross-power spectra are defined as
\begin{equation}
	\mathcal{P}_{ab}(k)
	\equiv
	\frac{k^3}{2\pi^2}
	\mathrm{Re}
	\left[
	\delta X_a(k)\, \delta X_b^*(k)
	\right].
	\label{power_spectra_definition}
\end{equation}
By comparing the two expressions for the shape function, the following general relation for $f_{\rm NL}$ is obtained:
\begin{equation}
	f_{\text{NL}} = \frac{5}{6}\frac{k_3^3 N_a N_b N_{cd} \mathcal{P}_{ac}(k_1) \mathcal{P}_{bd}(k_2) + \text{2 perms}}{k_3^3 N_e N_f N_g N_h \mathcal{P}_{ef}(k_1) \mathcal{P}_{gh}(k_2) + \text{2 perms}}.
	\label{f_NL_function_dN}
\end{equation}
Equations~\eqref{Shape_function_dN} and \eqref{f_NL_function_dN} can be used to study the shape and amplitude of non-Gaussianity in various configurations. In this work, we limit ourselves to the equilateral configuration ($k_1=k_2=k_3=k$) for simplicity and for comparison with existing literature. While the local shape is typically associated with multi-field models or superhorizon evolution, the gradient corrections introduce derivative couplings that naturally source intrinsic equilateral-type non-Gaussianities, peaking when modes have comparable wavelengths. In this template, one obtains
\begin{equation}
	f_{\rm NL}^{\rm eq}
	=
	\frac{5}{6}
	\frac{
		N_{ab} N_c N_d
		\mathcal{P}_{ac} (k)\mathcal{P}_{bd} (k)
	}{
		\left(
		N_e N_f \mathcal{P}_{ef} (k)
		\right)^2
	}.
	\label{f_NL_formula}
\end{equation}

In principle, one may incorporate gradient corrections into the source term order by order and compute $f_{\rm NL}^{\rm eq}$ from Eq.~\eqref{f_NL_formula} at each step, analogous to the power-spectrum analysis shown in Fig.~\ref{fig:compare_HM_deltaN_bump}. In practice, to facilitate a direct comparison with previous work, we numerically evaluate the source term using the linear mode functions obtained from the exact Mukhanov--Sasaki equation. This procedure allows the finite-$k$ dependence of the linear gradient correction to be fully incorporated into the $\delta N$ framework. Its main advantage is efficiency: nonlinear observables can then be estimated by solving the background dynamics together with the linear perturbation equations.

One could incorporate gradient corrections order by order into the source term and compute $f_{\rm NL}^{\rm eq}$ using Eq.~\eqref{f_NL_formula} at each step, analogous to the power spectrum analysis shown in Figure~\ref{fig:compare_HM_deltaN_bump}. However, to facilitate a direct and accurate comparison with previous works, we instead numerically evaluate the source term with full gradients derived from the exact MS equation. The resulting power spectrum is shown in Figure~\ref{fig:compare_HM_deltaN_bump}. The main advantage of this approach is its efficiency: nonlinear observables can be estimated by solving the background dynamics forced by the linear perturbation equations.

\begin{figure}[tbp]
	\centering
	\includegraphics[width=0.49\textwidth]{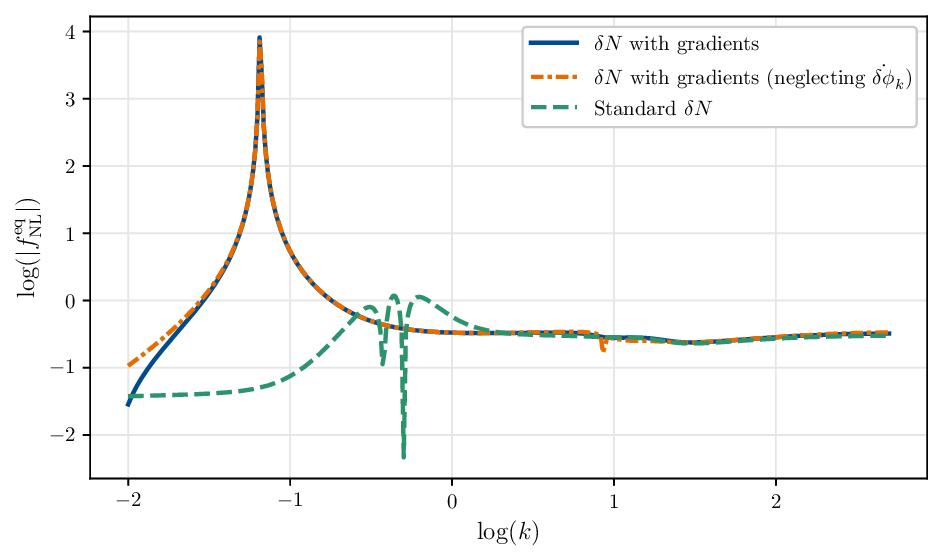}
	\hfill
	\includegraphics[width=0.49\textwidth]{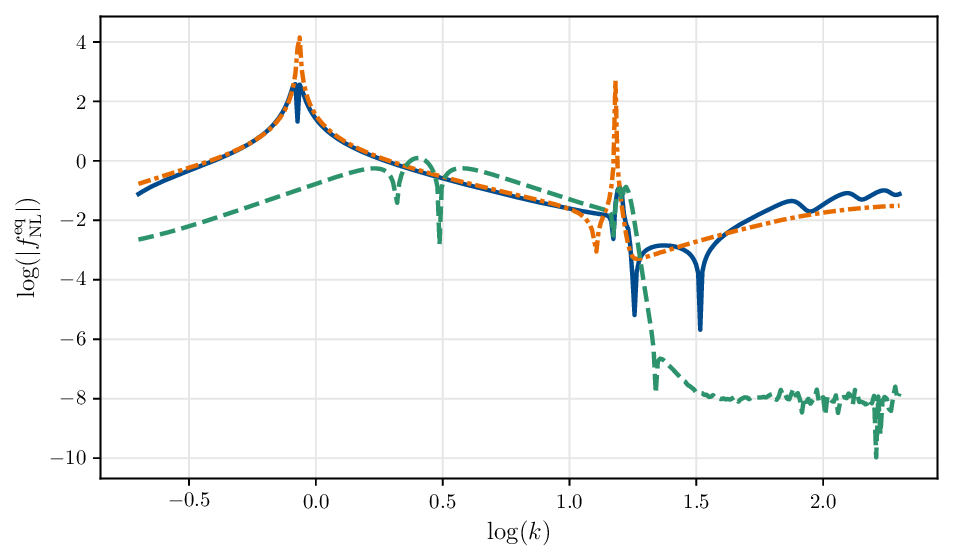}
	\caption{Equilateral-type non-Gaussianity parameter $f_{\rm NL}^{\rm eq}$ computed with gradient interactions (solid blue line) and using the standard $\delta N$ formalism (dashed green line) for the Gaussian bump (left panel) and Starobinsky (right panel) models \cite{github_code}. We use the same parameter values as in Figure \ref{fig:compare_HM_deltaN_bump}. The orange dot--dashed line shows the gradient-corrected $f_{\rm NL}^{\rm eq}$ obtained by neglecting the contribution of $\dot{\delta\phi}_k$. In all panels, the matching time is chosen at horizon crossing, $\sigma = 1$. For ease of comparison, the parameter values of the Starobinsky model are chosen similar to Ref.~\cite{hazra2013bingo}.}
	\label{fig:f_NL}
\end{figure}

Throughout this numerical calculation, all derivatives of $N$ in Eq.~\eqref{f_NL_formula} are evaluated once Eq.~\eqref{KG_with_correction} is solved. Using the relation $\delta \phi_k = \dot{\phi}\mathcal{R}_k/H$ and its derivative, we construct the power spectra defined in Eq.~\eqref{power_spectra_definition}. The resulting $f_{\rm NL}^{\rm eq}$ for the models introduced in the previous subsection are obtained numerically and shown in Figure~\ref{fig:f_NL}. The results indicate that $f_{\rm NL}^{\rm eq}$ peaks at scales where the power spectrum exhibits a pronounced dip, in excellent agreement with the perturbation theory results evaluated in Ref.~\cite{hazra2013bingo}. A comparison with the results obtained using the standard $\delta N$ formalism, also shown in Figure~\ref{fig:f_NL}, reveals that the standard approach fails to capture the dominant non-Gaussian features sourced by gradient interactions.

We note that the individual terms in Eq.~\eqref{f_NL_formula} can be positive or negative, leading to significant cancellations among seemingly dominant contributions. Consequently, subdominant terms may become numerically important. We also evaluated the gradient-corrected $f_{\rm NL}^{\rm eq}$ while neglecting contributions from $\dot{\delta \phi}_k$, as shown in Figure~\ref{fig:f_NL}. In the Starobinsky model, this approximation fails to capture certain non-Gaussian features, indicating that such simplifications should be adopted with caution.

Note that in Subsection~\ref{sec:the_source_term} we showed that the gradient corrected $\delta N$ formalism recover perturbation theory results at linear order. The authors of Ref.~\cite{cruces2023update} discuss this equivalence for nonlinear terms and conclude that the $\delta N$ formalism is only valid in perturbative regimes. In this subsection, we assumed that linear-order non-local interactions are more significant than higher-order local ones, and we justified this assumption by comparing the full numerical results obtained from our $\delta N$ formalism with those derived using the in-in formalism in Ref.~\cite{hazra2013bingo}.

\section{Conclusions and Outlook}
\label{sec:Summary}

In this work, we developed an extension of the $\delta N$ formalism based on a consistent and organized treatment of spatial gradient effects. The central result of this paper is the demonstration that the leading gradient-sensitive contributions to super-Hubble curvature perturbations can be captured by introducing an effective, gauge-invariant source term into the background Klein--Gordon equation. This construction provides a transparent and computationally efficient framework for tracking the evolution of perturbations from horizon exit to the end of inflation, even in regimes where the standard $\delta N$ formalism fails.

We showed that the source-term formulation is formally equivalent, at linear order, to the higher-order matching method based on the gradient expansion of the Mukhanov--Sasaki equation. Specifically, including $i$-th order gradient corrections through the source term reproduces the $(i+1)$-th order contribution in the gradient-expanded solution for the curvature perturbation. This establishes a direct and rigorous correspondence between gradient-corrected $\delta N$ dynamics and full linear perturbation theory, while preserving the intrinsically nonlinear character of the $\delta N$ approach.

Using two representative inflationary models that exhibit transient ultra-slow-roll phases (a smooth Gaussian-bump potential and the Starobinsky model with a sharp transition) we demonstrated that the gradient-corrected $\delta N$ formalism accurately reproduces the power spectrum obtained from the exact Mukhanov--Sasaki equation. Importantly, this agreement holds even when the matching between quantum and classical evolution is performed at horizon crossing, $\sigma = 1$, a regime where the standard separate universe approximation is known to break down. Our results also show that both adiabatic and non-adiabatic gradient corrections are essential for capturing the correct super-Hubble evolution in models relevant for primordial black hole formation.

Building on this validation, we applied the formalism to the computation of intrinsic non-Gaussianity by evaluating the equilateral bispectrum amplitude $f_{\mathrm{NL}}^{\mathrm{eq}}$. We showed that the non-Gaussian signal obtained using the gradient-corrected $\delta N$ formalism is in excellent agreement with perturbative in-in calculations, whereas the standard $\delta N$ formalism fails to capture these effects. We further demonstrated that neglecting contributions from seemingly subdominant initial conditions can lead to qualitatively incorrect results, underscoring the importance of retaining the full phase-space structure in gradient-corrected analyses.

We showed that the contribution of spatial gradients to the dynamics of the Mukhanov--Sasaki variable scales as $a^{-2}$, which is the same scaling as that of a spatial curvature term, $\mathcal{K}$, in the local universe approximation considered in \cite{artigas2025extended}. This pioneering work aims to extend the $\delta N$ formalism by incorporating spatial gradient effects through an effective spatial curvature assigned to each locally homogeneous FLRW patch. By explicitly comparing the resulting equations of motion in a curved FLRW background with those obtained from our source-term formulation, we showed that the curvature-based approach corresponds to an effective source term that does not fully account for $\mathcal{O}(k^2)$ adiabatic gradient correction.

The source-term formalism presented here provides a practical and powerful bridge between full cosmological perturbation theory and nonlinear separate-universe techniques. Within the present implementation, the source term $\mathcal{S}_k$ is computed under the approximation that $\mathcal{R}_k$ evolves according to first-order perturbation theory. Within this approximation, the backreaction effects of fluctuations on the background are neglected, which helps to make the problem tractable. A more consistent treatment would employ a recursive scheme in which the background and the source term are evaluated at the same order; the study of gradient interactions beyond the present approximation is therefore left for future work. Another promising direction is the application of this framework to the study of gradient-induced modifications of the probability distribution function of curvature perturbations. These results can then be compared with the stochastic $\delta N$ formalism, specifically in cosmological models with slow-roll violation, where noise plays a crucial role.

\section*{Data Availability Statement}
The codes required to reproduce the results presented in this work are freely available at GitHub~\cite{github_code}.

\acknowledgments
We acknowledge the financial support of the research council of the University of Tehran. We also thank Fatemeh Eghbalpoor for helpful discussions during the early stages of this work. Finally, we thank the anonymous referee for thoughtful comments and suggestions that substantially improved the quality and clarity of the manuscript.

\appendix

\section{\boldmath $\delta N$ formalism with spatial curvature}
\label{app:review_extended_deltaN}

In this appendix, we briefly review the $\delta N$ formalism extended to include spatial curvature in separate universes discussed in Ref.~\cite{artigas2025extended}. Our aim is not to reproduce the full derivation, but to summarize the main assumptions and equations. The key idea is to incorporate leading gradient effects by approximating each super-Hubble patch as a homogeneous FLRW universe with a local spatial curvature $\mathcal K$. The background equations for an FLRW metric with non-zero spatial curvature are
\begin{align}
	 \ddot{\phi} + 3 H \dot{\phi} + V_{,\phi} &= 0 \, ,
	\label{eq:bg_review_1}
	\\
	3 H^2 + 3 \frac{a_*^2}{a^2}  \mathcal K &= \frac{1}{2}\dot{\phi}^2 + V \, .
	\label{eq:bg_review_2}
\end{align}
The curvature contribution appearing in the Friedmann equation should be chosen properly such that it mimics the effect of gradients. To do so, the perturbed background equations and perturbation equations are compared. 

Perturbing the background equations yields
\begin{align}
	\ddot{\delta\phi}_k + 3H \dot{\delta\phi}_k + V_{,\phi\phi}\delta\phi_k - 2 \frac{\delta H_k}{H} V_{,\phi} + \dot{\phi} \frac{d}{dt}\left(\frac{\delta H_k}{H}\right) &= 0 \, ,
	\label{pert_BG_1}
	\\
	V_{,\phi} \delta\phi_k + \dot{\phi}\dot{\delta\phi}_k - 2 \frac{\delta H_k}{H} V + 3 \frac{a_*^2}{a^2}  \mathcal K &=0\, .
	\label{pert_BG_2}
\end{align}
On the other hand, the linearized KG equation \eqref{eq:kg_perturbed} and the energy constraint \eqref{eq:energy_constraint} in the uniform expansion gauge (in which $\delta N = B=0$) take the form:
\begin{align}
	\ddot{\delta\phi}_k + 3H \dot{\delta\phi}_k + V_{,\phi\phi}\delta\phi_k + 2A_k V_{,\phi} - \dot{A}_k\dot{\phi} + \frac{k^2}{a^2}\delta\phi_k &= 0 \, ,
	\label{eq:KG_pert_review}
	\\
	V_{,\phi} \delta\phi_k + \dot{\phi}\dot{\delta\phi}_k + 2 A_k V + 2 \frac{k^2 }{a^2} \psi_k &=0\, .
	\label{eq:A_review}
\end{align}
By comparing these equations with their counterparts, \eqref{pert_BG_1} and \eqref{pert_BG_2}, one notices that the perturbed background Friedmann equation coincides with the energy constraint~\eqref{eq:A_review} upon identifying
\begin{equation}
	\frac{\delta H_k}{H} = - A_k,
	\qquad
	\mathcal K = \frac{2k^2}{3 a_*^2}\psi_{k*} \, .
	\label{eq:K_identification}
\end{equation}
However, the contribution from gradient interactions, $k^2 \delta \phi / a^2$, is absent from the perturbed background KG equation~\eqref{pert_BG_1}.

In this scenario, the local dynamics depend on three quantities,
$
\{\phi_{\mathrm{in}}, \dot{\phi}_{\mathrm{in}},\, \mathcal K\},
$
but only two are physically independent. Let us use the residual gauge freedom present in the uniform expansion gauge to set 
$
\delta\phi_{k*} = 0
$. This leaves
$
X = (\dot{\phi}_{\mathrm{in}},\, \mathcal K)
$
as the independent parameters. This mirrors the situation discussed in the main text, where four integration constants reduce to two independent degrees of freedom in single-clock inflation. Using the constraint equations, Eq.~\eqref{eq:energy_constraint} and Eq.~\eqref{eq:momentum_constraint}, the initial conditions for other perturbations can be written in terms of the gauge-invariant quantity $\mathcal{R}_k$ at the matching time:
\begin{equation}
	 \psi_{k*} = \mathcal{R}_{k*}, \quad \dot{\delta\phi_{k*}} = \frac{\dot{\phi}_*}{3 H^2} \left[ \frac{V }{a_* H} \mathcal{R}'_{k*} - \frac{k^2}{a_*^2} \mathcal{R}_{k*} \right].
\end{equation}

\begin{figure}[tbp]
	\centering
	\includegraphics[width=0.49\textwidth]{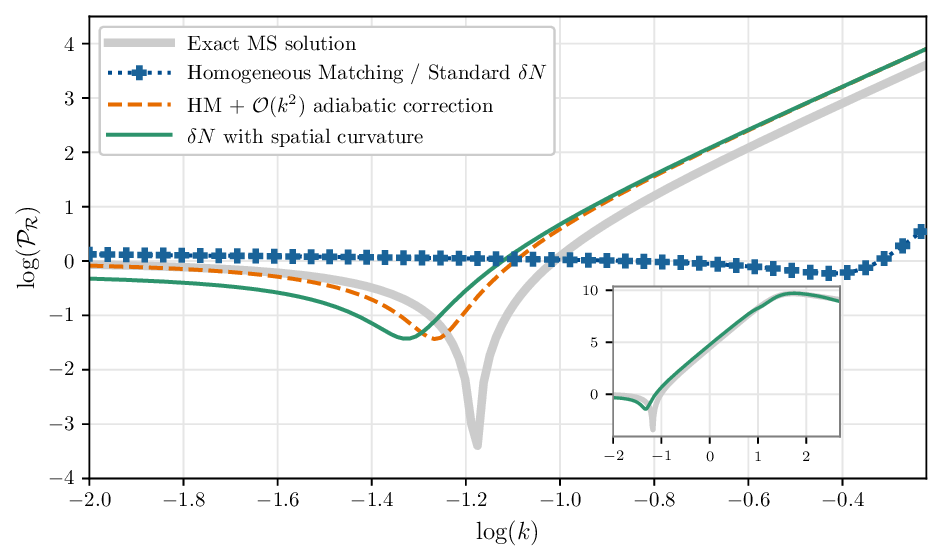}
	\includegraphics[width=0.49\textwidth]{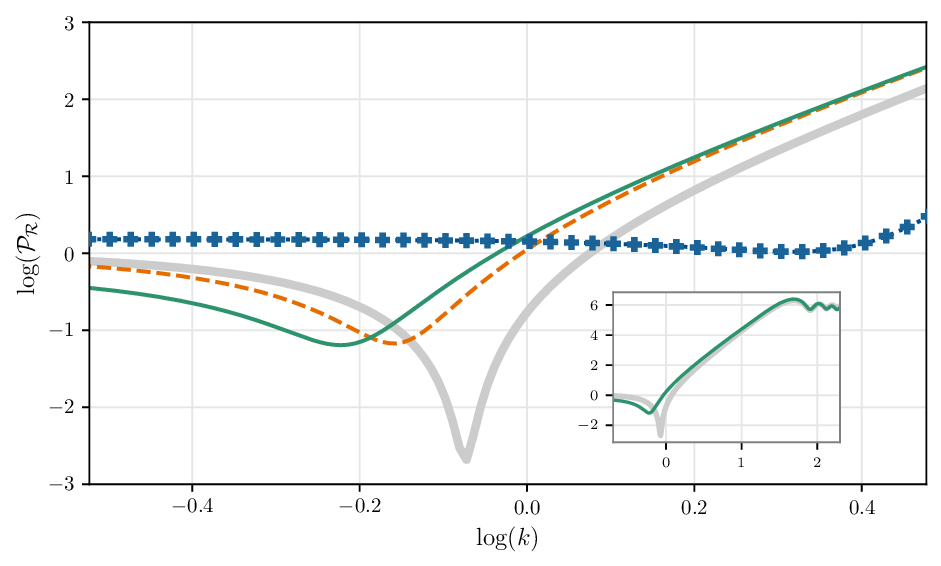}
	\caption{Comparison of the power spectra obtained from the exact MS equation (gray line), the standard $\delta N$ formalism (blue markers), the $\delta N$ formalism with spatial curvature (green line), the homogeneous matching method (dotted blue line), and the higher-order matching method with the leading adiabatic correction (orange dashed line), for the Gaussian bump (left panel) and Starobinsky (right panel) models \cite{github_code}. The parameter values are chosen similar to those in \ref{fig:compare_HM_deltaN_bump}. The inset panels compare the MS solution with the $\delta N$ formalism with spatial curvature over a wider range of wavenumbers.
	}
	\label{fig:dN_with_kappa}
\end{figure}

We have evaluated the power spectrum using this $\delta N$ formalism with spatial curvature for the two models discussed in the main text, using the full numerical procedure. The results are shown in Figure~\ref{fig:dN_with_kappa}. It is clear that the inclusion of spatial curvature is a significant improvement compared to the standard $\delta N$ formalism, yet the resulting power spectrum does not accurately reproduce the leading order adiabatic $k^2$ correction.

It is common to apply gradient interactions to part of the background equations, working in the decoupling limit, for example \cite{briaud2025stochastic}, but its justification should be explicit. A comparison of this approach with our approach in this work illustrates why this discrepancy arises. For this purpose, we compare the effective equations of motion for the scalar field in both methods. Incorporating the spatial curvature into the field dynamics leads to a background evolution equation of the form:
\begin{equation}
	\phi_{,nn} + \left(3 - \frac{1}{2}\phi_{,n}^2\right)
	\left(
	\phi_{,n}+\frac{V_{,\phi} +   a_*^2 \mathcal{K} \phi_{,n}/ a^2 }{V - 3 a_*^2 \mathcal K / a^2}\right)
	= 0 \, ,
\end{equation}
while for the method with the source term defined in Eq.~\eqref{KG_with_correction}, we get
\begin{equation}
	\phi_{,nn} + \left(3 - \frac{1}{2}\phi_{,n}^2\right)
	\left(
	\phi_{,n} +\frac{V_{,\phi} - \mathcal{S}_k}{V}\right)
	= 0 \,.
\end{equation}
Expanding in terms of $\mathcal{K}$ up to first order and comparing the two equations, it appears that the spatial curvature method is analogous to using a source term, $\tilde{\mathcal{S}}_k$, given by
\begin{equation}
	\tilde{\mathcal{S}}_k = \left(2 + \frac{3 \eta }{6-2 \epsilon }\right) \frac{a_*^2 \dot{\phi}}{a^2 H} \mathcal{K} = \left(\frac{4}{3}-\frac{\eta }{\epsilon -3}\right) \frac{k^2 \dot{\phi}}{a^2 H} \mathcal{R}_{k*}.
	\label{identification_of_source_term}
\end{equation}
This expression should be compared with
$
\mathcal{S}_k = - \frac{k^2 \dot{\phi}}{a^2 H}\,\mathcal{R}_{k*} \, ,
$
which is known to accurately reproduce the leading adiabatic gradient correction. We observe that the two source terms differ by a time-dependent factor proportional to $\eta$. Therefore, in non-attractor models of inflation with large $\eta$, which are typically of interest in PBH studies, we expect the resulting gradient-corrected power spectrum derived from this formalism to differ slightly from that obtained using the leading-order adiabatic gradient expansion. This slight discrepancy is evident in the diagrams shown in Figure~\ref{fig:dN_with_kappa}.

\bibliographystyle{JHEP}
\bibliography{dNwithgradients}

@misc{github_code,
  howpublished = {The Python codes used to plot all figures in this work are publicly available at \url{https://github.com/mohammadahmadi-physics/dN-formalism-with-gradient-interactions}},
  
}

@article{starobinsky1980new,
  title={A new type of isotropic cosmological models without singularity},
  author={Starobinsky, Alexei A},
  journal={Physics Letters B},
  volume={91},
  number={1},
  pages={99--102},
  year={1980},
  publisher={Elsevier}
}

@article{sato1981first,
  title={First-order phase transition of a vacuum and the expansion of the Universe},
  author={Sato, Katsuhiko},
  journal={Monthly Notices of the Royal Astronomical Society},
  volume={195},
  number={3},
  pages={467--479},
  year={1981},
  publisher={Oxford University Press Oxford, UK}
}

@article{guth1981inflationary,
  title={Inflationary universe: A possible solution to the horizon and flatness problems},
  author={Guth, Alan H},
  journal={Physical Review D},
  volume={23},
  number={2},
  pages={347},
  year={1981},
  publisher={APS}
}

@article{albrecht1982reheating,
  title={Reheating an inflationary universe},
  author={Albrecht, Andreas and Steinhardt, Paul J and Turner, Michael S and Wilczek, Frank},
  journal={Physical Review Letters},
  volume={48},
  number={20},
  pages={1437},
  year={1982},
  publisher={APS}
}

@article{linde1982new,
  title={A new inflationary universe scenario: a possible solution of the horizon, flatness, homogeneity, isotropy and primordial monopole problems},
  author={Linde, Andrei D},
  journal={Physics Letters B},
  volume={108},
  number={6},
  pages={389--393},
  year={1982},
  publisher={Elsevier}
}

@article{lyth1999particle,
  title={Particle physics models of inflation and the cosmological density perturbation},
  author={Lyth, David H and Riotto, Antonio},
  journal={Physics Reports},
  volume={314},
  number={1-2},
  pages={1--146},
  year={1999},
  publisher={Elsevier}
}

@article{artigas2025extended,
  title={Extended {$\delta N$} Formalism: Nonspatially Flat Separate-Universe Approach},
  author={Artigas, Danilo and Pi, Shi and Tanaka, Takahiro},
  journal={Physical Review Letters},
  volume={134},
  number={22},
  pages={221001},
  year={2025},
  publisher={APS}
}

@article{briaud2025stochastic,
  title={Stochastic inflation with gradient interactions},
  author={Briaud, Vadim and Kawaguchi, Ryodai and Vennin, Vincent},
  journal={Journal of Cosmology and Astroparticle Physics},
  volume={2025},
  number={12},
  pages={024},
  year={2025},
  publisher={IOP Publishing}
}

@article{jackson2024separate,
  title={The separate-universe approach and sudden transitions during inflation},
  author={Jackson, Joseph HP and Assadullahi, Hooshyar and Gow, Andrew D and Koyama, Kazuya and Vennin, Vincent and Wands, David},
  journal={Journal of Cosmology and Astroparticle Physics},
  volume={2024},
  number={05},
  pages={053},
  year={2024},
  publisher={IOP Publishing}
}

@article{wands2000new,
  title={New approach to the evolution of cosmological perturbations on large scales},
  author={Wands, David and Malik, Karim A and Lyth, David H and Liddle, Andrew R},
  journal={Physical Review D},
  volume={62},
  number={4},
  pages={043527},
  year={2000},
  publisher={APS}
}

@article{lyth2003conserved,
  title={Conserved cosmological perturbations},
  author={Lyth, David H and Wands, David},
  journal={Physical Review D},
  volume={68},
  number={10},
  pages={103515},
  year={2003},
  publisher={APS}
}

@article{rigopoulos2003separate,
  title={Separate universe approach and the evolution of nonlinear superhorizon cosmological perturbations},
  author={Rigopoulos, GI and Shellard, EPS},
  journal={Physical Review D},
  volume={68},
  number={12},
  pages={123518},
  year={2003},
  publisher={APS}
}

@article{pattison2019stochastic,
  title={Stochastic inflation beyond slow roll},
  author={Pattison, Chris and Vennin, Vincent and Assadullahi, Hooshyar and Wands, David},
  journal={Journal of Cosmology and Astroparticle Physics},
  volume={2019},
  number={07},
  pages={031},
  year={2019},
  publisher={IOP Publishing}
}

@article{starobinsky1982dynamics,
  title={Dynamics of phase transition in the new inflationary universe scenario and generation of perturbations},
  author={Starobinsky, Alexei A},
  journal={Physics Letters B},
  volume={117},
  number={3-4},
  pages={175--178},
  year={1982},
  publisher={Elsevier}
}

@article{starobinskiǐ1985multicomponent,
  title={Multicomponent de Sitter (inflationary) stages and the generation of perturbations},
  author={Starobinskiǐ, AA},
  journal={Soviet Journal of Experimental and Theoretical Physics Letters},
  volume={42},
  pages={152},
  year={1985}
}

@article{sasaki1996general,
  title={A General analytic formula for the spectral index of the density perturbations produced during inflation},
  author={Sasaki, Misao and Stewart, Ewan D},
  journal={Progress of Theoretical Physics},
  volume={95},
  number={1},
  pages={71--78},
  year={1996},
  publisher={Oxford University Press}
}

@article{sasaki1998super,
  title={Super-horizon scale dynamics of multi-scalar inflation},
  author={Sasaki, Misao and Tanaka, Takahiro},
  journal={Progress of Theoretical Physics},
  volume={99},
  number={5},
  pages={763--781},
  year={1998},
  publisher={Oxford University Press}
}

@article{lyth2005general,
  title={A general proof of the conservation of the curvature perturbation},
  author={Lyth, David H and Malik, Karim A and Sasaki, Misao},
  journal={Journal of Cosmology and Astroparticle Physics},
  volume={2005},
  number={05},
  pages={004},
  year={2005},
  publisher={IOP Publishing}
}

@article{yokoyama2008primordial,
  title={Primordial non-Gaussianity in multiscalar inflation},
  author={Yokoyama, Shuichiro and Suyama, Teruaki and Tanaka, Takahiro},
  journal={Physical Review D—Particles, Fields, Gravitation, and Cosmology},
  volume={77},
  number={8},
  pages={083511},
  year={2008},
  publisher={APS}
}

@article{wands2010local,
  title={Local non-Gaussianity from inflation},
  author={Wands, David},
  journal={Classical and Quantum Gravity},
  volume={27},
  number={12},
  pages={124002},
  year={2010},
  publisher={IOP Publishing}
}

@article{hooshangi2022rare,
  title={Rare events are nonperturbative: Primordial black holes from heavy-tailed distributions},
  author={Hooshangi, Sina and Namjoo, Mohammad Hossein and Noorbala, Mahdiyar},
  journal={Physics Letters B},
  volume={834},
  pages={137400},
  year={2022},
  publisher={Elsevier}
}

@article{hooshangi2023tail,
  title={Tail diversity from inflation},
  author={Hooshangi, Sina and Namjoo, Mohammad Hossein and Noorbala, Mahdiyar},
  journal={Journal of Cosmology and Astroparticle Physics},
  volume={2023},
  number={09},
  pages={023},
  year={2023},
  publisher={IOP Publishing}
}

@article{mishra2020primordial,
  title={Primordial Black Holes from a tiny bump/dip in the Inflaton potential},
  author={Mishra, Swagat S and Sahni, Varun},
  journal={Journal of Cosmology and Astroparticle Physics},
  volume={2020},
  number={04},
  pages={007},
  year={2020},
  publisher={IOP Publishing}
}

@article{cole2023primordial,
  title={Primordial black holes from single-field inflation: a fine-tuning audit},
  author={Cole, Philippa S and Gow, Andrew D and Byrnes, Christian T and Patil, Subodh P},
  journal={Journal of Cosmology and Astroparticle Physics},
  volume={2023},
  number={08},
  pages={031},
  year={2023},
  publisher={IOP Publishing}
}

@article{starobinskij1992spectrum,
  title={Spectrum of adiabatic perturbations in the universe when there are singularities in the inflationary potential.},
  author={Starobinskij, AA},
  journal={Soviet Journal of Experimental and Theoretical Physics Letters},
  volume={55},
  number={9},
  pages={489--494},
  year={1992}
}

@article{martin2012scalar,
  title={The scalar bi-spectrum in the Starobinsky model: The equilateral case},
  author={Martin, Jerome and Sriramkumar, L},
  journal={Journal of Cosmology and Astroparticle Physics},
  volume={2012},
  number={01},
  pages={008},
  year={2012},
  publisher={IOP Publishing}
}

@article{martin2014sharp,
  title={Sharp inflaton potentials and bi-spectra: Effects of smoothening the discontinuity},
  author={Martin, Jerome and Sriramkumar, L and Hazra, Dhiraj Kumar},
  journal={Journal of Cosmology and Astroparticle Physics},
  volume={2014},
  number={09},
  pages={039},
  year={2014},
  publisher={IOP Publishing}
}

@article{ahmadi2022quantum,
  title={Quantum diffusion in sharp transition to non-slow-roll phase},
  author={Ahmadi, Nahid and Noorbala, Mahdiyar and Feyzabadi, Niloufar and Eghbalpoor, Fatemeh and Ahmadi, Zahra},
  journal={Journal of Cosmology and Astroparticle Physics},
  volume={2022},
  number={08},
  pages={078},
  year={2022},
  publisher={IOP Publishing}
}

@article{zel1966hypothesis,
  title={The hypothesis of cores retarded during expansion and the hot cosmological model},
  author={Zel'dovich, Ya B and Novikov, ID},
  journal={Astronomicheskii Zhurnal},
  volume={43},
  pages={758},
  year={1966}
}

@article{hawking1971gravitationally,
  title={Gravitationally collapsed objects of very low mass},
  author={Hawking, Stephen},
  journal={Monthly Notices of the Royal Astronomical Society},
  volume={152},
  number={1},
  pages={75--78},
  year={1971},
  publisher={Oxford University Press}
}

@article{carr1974black,
  title={Black holes in the early Universe},
  author={Carr, Bernard J and Hawking, Stephen W},
  journal={Monthly Notices of the Royal Astronomical Society},
  volume={168},
  number={2},
  pages={399--415},
  year={1974},
  publisher={Oxford University Press Oxford, UK}
}

@article{carr1975primordial,
  title={The Primordial black hole mass spectrum},
  author={Carr, Bernard J},
  journal={Astrophysical Journal, vol. 201, Oct. 1, 1975, pt. 1, p. 1-19. Research supported by the Science Research Council of England},
  volume={201},
  pages={1--19},
  year={1975}
}

@article{chapline1975cosmological,
  title={Cosmological effects of primordial black holes},
  author={Chapline, George F},
  journal={Nature},
  volume={253},
  number={5489},
  pages={251--252},
  year={1975},
  publisher={Nature Publishing Group UK London}
}

@article{carr1984can,
  title={Can pregalactic objects generate galaxies?},
  author={Carr, BJ and Rees, Martin J},
  journal={Monthly Notices of the Royal Astronomical Society},
  volume={206},
  number={4},
  pages={801--818},
  year={1984},
  publisher={Oxford University Press Oxford, UK}
}

@article{byrnes2012primordial,
  title={Primordial black holes as a tool for constraining non-Gaussianity},
  author={Byrnes, Christian T and Copeland, Edmund J and Green, Anne M},
  journal={Physical Review D—Particles, Fields, Gravitation, and Cosmology},
  volume={86},
  number={4},
  pages={043512},
  year={2012},
  publisher={APS}
}

@article{young2013primordial,
  title={Primordial black holes in non-Gaussian regimes},
  author={Young, Sam and Byrnes, Christian T},
  journal={Journal of Cosmology and Astroparticle Physics},
  volume={2013},
  number={08},
  pages={052},
  year={2013},
  publisher={IOP Publishing}
}

@article{passaglia2019primordial,
  title={Primordial black holes and local non-Gaussianity in canonical inflation},
  author={Passaglia, Samuel and Hu, Wayne and Motohashi, Hayato},
  journal={Physical Review D},
  volume={99},
  number={4},
  pages={043536},
  year={2019},
  publisher={APS}
}

@article{atal2019role,
  title={The role of non-gaussianities in Primordial Black Hole formation},
  author={Atal, Vicente and Germani, Cristiano},
  journal={Physics of the Dark Universe},
  volume={24},
  pages={100275},
  year={2019},
  publisher={Elsevier}
}

@article{biagetti2021formation,
  title={The formation probability of primordial black holes},
  author={Biagetti, Matteo and De Luca, Valerio and Franciolini, Gabriele and Kehagias, Alex and Riotto, Antonio},
  journal={Physics Letters B},
  volume={820},
  pages={136602},
  year={2021},
  publisher={Elsevier}
}

@article{taoso2021non,
  title={Non-gaussianities for primordial black hole formation},
  author={Taoso, Marco and Urbano, Alfredo},
  journal={Journal of Cosmology and Astroparticle Physics},
  volume={2021},
  number={08},
  pages={016},
  year={2021},
  publisher={IOP Publishing}
}

@article{young2022peaks,
  title={Peaks and primordial black holes: the effect of non-Gaussianity},
  author={Young, Sam},
  journal={Journal of Cosmology and Astroparticle Physics},
  volume={2022},
  number={05},
  pages={037},
  year={2022},
  publisher={IOP Publishing}
}

@article{ferrante2022primordial,
  title={Primordial non-Gaussianity up to all orders: Theoretical aspects and implications for primordial black hole models},
  author={Ferrante, Giacomo and Franciolini, Gabriele and Iovino, Antonio Junior and Urbano, Alfredo},
  journal={arXiv preprint arXiv:2211.01728},
  year={2022}
}

@article{gow2023non,
  title={Non-perturbative non-Gaussianity and primordial black holes},
  author={Gow, Andrew D and Assadullahi, Hooshyar and Jackson, Joseph HP and Koyama, Kazuya and Vennin, Vincent and Wands, David},
  journal={Europhysics Letters},
  volume={142},
  number={4},
  pages={49001},
  year={2023},
  publisher={IOP Publishing}
}

@article{komatsu2001acoustic,
  title={Acoustic signatures in the primary microwave background bispectrum},
  author={Komatsu, Eiichiro and Spergel, David N},
  journal={Physical Review D},
  volume={63},
  number={6},
  pages={063002},
  year={2001},
  publisher={APS}
}

@article{maldacena2003non,
  title={Non-Gaussian features of primordial fluctuations in single field inflationary models},
  author={Maldacena, Juan},
  journal={Journal of High Energy Physics},
  volume={2003},
  number={05},
  pages={013},
  year={2003},
  publisher={IOP Publishing}
}

@article{bartolo2004non,
  title={Non-Gaussianity from inflation: Theory and observations},
  author={Bartolo, Nicola and Komatsu, Eiichiro and Matarrese, Sabino and Riotto, Antonio},
  journal={Physics Reports},
  volume={402},
  number={3-4},
  pages={103--266},
  year={2004},
  publisher={Elsevier}
}

@article{liguori2010primordial,
  title={Primordial Non-Gaussianity and Bispectrum Measurements in the Cosmic Microwave Background and Large-Scale Structure},
  author={Liguori, Michele and Sefusatti, Emiliano and Fergusson, James R and Shellard, EPS},
  journal={Advances in Astronomy},
  volume={2010},
  number={1},
  pages={980523},
  year={2010},
  publisher={Wiley Online Library}
}

@article{chen2010primordial,
  title={Primordial Non-Gaussianities from Inflation Models},
  author={Chen, Xingang},
  journal={Advances in Astronomy},
  volume={2010},
  number={1},
  pages={638979},
  year={2010},
  publisher={Wiley Online Library}
}

@article{wang2014inflation,
  title={Inflation, cosmic perturbations and non-Gaussianities},
  author={Wang, Yi},
  journal={Communications in Theoretical Physics},
  volume={62},
  number={1},
  pages={109},
  year={2014},
  publisher={IOP Publishing}
}

@article{artigas2022hamiltonian,
  title={Hamiltonian formalism for cosmological perturbations: the separate-universe approach},
  author={Artigas, Danilo and Grain, Julien and Vennin, Vincent},
  journal={Journal of Cosmology and Astroparticle Physics},
  volume={2022},
  number={02},
  pages={001},
  year={2022},
  publisher={IOP Publishing}
}

@article{hazra2013bingo,
  title={BINGO: A code for the efficient computation of the scalar bi-spectrum},
  author={Hazra, Dhiraj Kumar and Sriramkumar, L and Martin, Jerome},
  journal={Journal of Cosmology and Astroparticle Physics},
  volume={2013},
  number={05},
  pages={026},
  year={2013},
  publisher={IOP Publishing}
}

@article{jain2008double,
  title={Punctuated inflation and the low CMB multipoles},
  author={Jain, Rajeev Kumar and Chingangbam, Pravabati and Gong, Jinn-Ouk and Sriramkumar, L and Souradeep, Tarun},
  journal={arXiv preprint arXiv:0809.3915},
  year={2008},
  publisher={Citeseer}
}

@article{jain2010tensor,
  title={Tensor-to-scalar ratio in punctuated inflation},
  author={Jain, Rajeev Kumar and Chingangbam, Pravabati and Sriramkumar, L and Souradeep, Tarun},
  journal={Physical Review D—Particles, Fields, Gravitation, and Cosmology},
  volume={82},
  number={2},
  pages={023509},
  year={2010},
  publisher={APS}
}

@article{namjoo2025geometry,
  title={Geometry of non-Gaussianity in transient non-attractor inflation},
  author={Namjoo, Mohammad Hossein and Nikbakht, Bahar},
  journal={arXiv preprint arXiv:2512.11020},
  year={2025}
}

@article{cai2018revisiting,
  title={Revisiting non-Gaussianity from non-attractor inflation models},
  author={Cai, Yi-Fu and Chen, Xingang and Namjoo, Mohammad Hossein and Sasaki, Misao and Wang, Dong-Gang and Wang, Ziwei},
  journal={Journal of Cosmology and Astroparticle Physics},
  volume={2018},
  number={05},
  pages={012},
  year={2018},
  publisher={IOP Publishing}
}

@article{cruces2023update,
  title={An update on adiabatic modes in cosmology and $\delta$ N formalism},
  author={Cruces, Diego and Germani, Cristiano and Palomares, Adrian},
  journal={Journal of Cosmology and Astroparticle Physics},
  volume={2023},
  number={06},
  pages={002},
  year={2023},
  publisher={IOP Publishing}
}

@article{kodama1998evolution,
  title={Evolution of cosmological perturbations in the long wavelength limit},
  author={Kodama, Hideo and Hamazaki, Takashi},
  journal={Physical Review D},
  volume={57},
  number={12},
  pages={7177},
  year={1998},
  publisher={APS}
}

@article{cruces2022review,
  title={Review on stochastic approach to inflation},
  author={Cruces, Diego},
  journal={Universe},
  volume={8},
  number={6},
  pages={334},
  year={2022},
  publisher={MDPI},
  eprint={arXiv:2203.13852}
}

@article{martin2013ultra,
  title={Ultra slow-roll inflation and the non-Gaussianity consistency relation},
  author={Martin, Jerome and Motohashi, Hayato and Suyama, Teruaki},
  journal={Physical Review D},
  volume={87},
  number={2},
  pages={023514},
  year={2013},
  publisher={APS},
  eprint={arXiv:1211.0083}
}

@article{motohashi2015inflation,
  title={Inflation with a constant rate of roll},
  author={Motohashi, Hayato and Starobinsky, Alexei A and Yokoyama, Jun'ichi},
  journal={Journal of Cosmology and Astroparticle Physics},
  number={09},
  pages={018},
  year={2015},
  publisher={IOP Publishing},
  eprint={arXiv:1411.5021}
}

@article{odintsov2017inflation,
  title={Inflation with a smooth constant-roll to constant-roll era transition},
  author={Odintsov, SD and Oikonomou, VK},
  journal={Physical Review D},
  volume={96},
  number={2},
  pages={024029},
  year={2017},
  publisher={APS}
}

@article{anguelova2018systematics,
  title={Systematics of constant roll inflation},
  author={Anguelova, Lilia and Suranyi, Peter and Wijewardhana, LC Rohana},
  journal={Journal of Cosmology and Astroparticle Physics},
  number={02},
  year={2018},
  publisher={IOP Publishing},
  eprintclass={hep-th},
  eprinttype={arxiv},
  eprint={arXiv:1710.06989}
}

@article{morse2018large,
  title={Large-$\eta$ constant-roll inflation is never an attractor},
  author={Morse, Michael JP and Kinney, William H},
  journal={Physical Review D},
  volume={97},
  number={12},
  year={2018},
  publisher={APS},
  eprinttype={arxiv},
  eprint={arXiv:1804.01927}
}

@article{yi2018constant,
  title={On the constant-roll inflation},
  author={Yi, Zhu and Gong, Yungui},
  journal={Journal of Cosmology and Astroparticle Physics},
  number={03},
  pages={052},
  year={2018},
  publisher={IOP Publishing},
  eprint={arXiv:1712.07478}
}

@article{lin2019dynamical,
  title={Dynamical analysis of attractor behavior in constant roll inflation},
  author={Lin, Wei-Chen and Morse, Michael JP and Kinney, William H},
  journal={Journal of Cosmology and Astroparticle Physics},
  number={09},
  year={2019},
  publisher={IOP Publishing},
  eprinttype={arxiv},
  eprint={arXiv:1904.06289}
}

@article{ghersi2019observational,
  title={Observational constraints on constant roll inflation},
  author={Ghersi, Jos{\'e} T G{\'a}lvez and Zucca, Alex and Frolov, Andrei V},
  journal={Journal of Cosmology and Astroparticle Physics},
  number={05},
  pages={030},
  year={2019},
  publisher={IOP Publishing},
  eprint={arXiv:1808.01325}
}

@article{mohammadi2023constant,
  title={On the constant roll complex scalar field inflationary models},
  author={Mohammadi, Ali and Ahmadi, Nahid and Shokri, Mehdi},
  journal={Journal of Cosmology and Astroparticle Physics},
  number={06},
  pages={058},
  year={2023},
  publisher={IOP Publishing},
  eprint={arXiv:2212.13403}
}

@article{mohammad2024analytical,
  title={Analytical insights into constant-roll condition: extending the paradigm to non-canonical models},
  author={Mohammad Ahmadi, S and Ahmadi, Nahid and Shokri, Mehdi},
  journal={Journal of Cosmology and Astroparticle Physics},
  volume={2024},
  number={05},
  pages={005},
  year={2024},
  publisher={IOP Publishing}
}

\end{document}